\documentclass[epj,nopacs]{svjour}
\usepackage{geometry}                
\usepackage{graphicx}
\usepackage{amssymb}
\usepackage{amsmath}

\begin{document}

\title{Trial factors for the look elsewhere effect in high energy physics}

\author{Eilam Gross \and Ofer Vitells}

\institute{Weizmann Institute of Science, Rehovot 76100, Israel}
\mail{ofer.vitells@weizmann.ac.il}


\abstract{ When searching for a new resonance somewhere in a
possible mass range, the significance of observing a local excess of
events must take into account the probability of observing such an
excess \emph{anywhere} in the range. This is the so called ``look
elsewhere effect''. The effect can be quantified in terms of a trial
factor, which is the ratio between the probability of observing the
excess at some fixed mass point, to the probability of observing it
anywhere in the range. We propose a simple and fast procedure for
estimating the trial factor, based on earlier results by Davies. We
show that asymptotically, the trial factor grows linearly with the
(fixed mass) significance. } \maketitle

\section{Introduction}
The statistical significance that is associated to the observation
of new phenomena is usually expressed using a \emph{p-value}, that
is, the probability that a similar or more extreme effect would be
seen when the signal does not exist (a situation usually referred to
as the null or background-only hypothesis). It is often the case
that one does not \emph{a priori} know where the signal will appear
within some possible range. In that case, the significance
calculation must take into account the fact that an excess of events
anywhere in the range could equally be considered as a signal. This
is known as the ``look elsewhere effect'' \cite{lyons}\cite{luc}. In
the statistical literature this situation is usually referred to as
an hypothesis test when a nuisance parameter is present only under
the alternative, for which the standard regularity conditions do not
apply. The problem is however closely related to that of a level
crossings of a stochastic process, which has been studied
extensively and for which many relevant results exist
\cite{davis77},\cite{davis87}. In particular \cite{davis87} provides
an upper bound on the tail probability of the maximum of a
chi-squared process, which can therefore be applied to cases when
the test statistic is the profile likelihood ratio and the
large-sample distribution at a fixed point given by Wilks' theorem
\cite{wilks} holds.

Of course, a straightforward way of quantifying the look-elsewhere
effect may be simply running many Monte-Carlo simulations of
background only experiments, and finding for each one the largest
fluctuation that resembles a signal\footnote{It is assumed that the
signal can appear only in one location, therefore when a search is
conducted, one will look for the largest excess of events and regard
all others as background fluctuations.}. Examples for such an
approach can be found e.g. in \cite{examples}. While this procedure
is simple and gives the correct answer, it is also time and CPU
consuming, as one would have to repeat it $\mathcal{O}(10^7)$ times
to get the p-value corresponding to a $5\sigma$ significance.
Approximations based on large sample limits such as \cite{davis87}
can therefore be valuable.

To make the discussion concrete, let us first describe a typical
example of a `mass bump' search. The statistical model in this case
consists of a background distribution $b$, and a signal distribution
$s(m)$ with an unknown location parameter $m$ (corresponding to the
mass of a resonance). We introduce a signal strength parameter
$\mu$, such that the model is given by $\mu s(m) + b$, and we test
$H_0:\mu=0$ against the alternative, $\mu>0$ \footnote{Strictly
speaking, Wilks' theorem does not apply here since the tested value
of $\mu$ is on the boundary of the allowed region. This difficulty
however can be easily alleviated by extending the allowed region of
$\mu$ to negative values and then reducing the p-value by half. A
formal generalization of this nature is given in \cite{chernoff}.}.
Both $s$ and $b$ may depend on additional nuisance parameters which
are omitted from the notation. The mass $m$ is another nuisance
parameter, but it does not exist under $H_0$ since $b$ does not
depend on $m$. The testing problem has therefore a nuisance
parameter that is present only under the alternative. When searching
for a resonance that can appear anywhere in the range, one will look
for the largest excess of events above the background. More
precisely, if $q(m)$ is a test statistic for a fixed mass, and large
values of $q(m)$ indicate decreasing compatibility with the null
hypothesis, then the test statistic of the entire range would be
$q(\hat m) = \displaystyle\max_m [q(m)] $. The problem is therefore
in assessing the tail probability (p-value) of the maximum of $q(m)$
over the search range.

In section \ref{sec1} we review the main result of \cite{davis87}
and its application to the present case. We propose a practical
procedure for estimating the p-value, which we then demonstrate with
a toy model simulation in section \ref{sec:toy}.

\section{Tail probabilities of the likelihood ratio test statistic}
\label{sec1} Suppose that a signal hypothesis depends on a nuisance
parameter $\theta$ that does not exist under the null. We denote by
$q(\theta)$ the profile likelihood test statistic for some fixed
$\theta$ and we assume that it follows a $\chi^2$ distribution with
$s$ degrees of freedom, as would be the case in the large sample
limit when Wilks' theorem holds. We are interested in the tail
probability of the maximum of $q(\theta)$ over $\theta$, which we
denote by $q(\hat\theta)$. As shown in \cite{davis87}, this is
bounded by:
\begin{equation}\label{eq1}
P(q(\hat\theta) > c) \leqslant P(\chi^2_s > c) + \langle N(c)
\rangle
\end{equation}

where $N(c)$ is the number of `upcrossings' of the level $c$ by the
process $q(\theta)$, with an expectation that is given by
\cite{davis87}:

\begin{equation}\label{eq2}
\langle N(c) \rangle = \frac{c^{(s-1)/2}e^{-c/2}}{
\sqrt{\pi}2^{s/2}\Gamma(s/2+1/2)} \int_L^U C(\theta) d\theta
\end{equation}

Where $[L,U]$ is the range of possible values of $\theta$, and
$C(\theta)$ is some function that depends on the details of the
statistical model.
 To have the maximum of $q(\theta)$ above the level $c$ means that
 either the value at the lower threshold $q(L)$ is larger than $c$,
 or that there is at least one upcrossing, hence the two terms in
 (\ref{eq1}). Note that the bound is expected to become an equality for large
values of $c$, as the expected number of upcrossings will be
dominated by the probability of one upcrossing, that is when
$\langle N(c) \rangle \ll 1$.

The function $C(\theta)$ can in general be difficult to calculate.
Instead, we propose to estimate $\langle N(c_0) \rangle$ at some low
reference level $c_0$ by simply counting the number of upcrossings
in a small set of background-only Monte Carlo simulations. Eq.
(\ref{eq1}) then becomes
\begin{equation}\label{eq3}
P(q(\hat\theta) > c) \leqslant P(\chi^2_s > c) + \langle N(c_0)
\rangle (\frac{c}{c_0})^{(s-1)/2} e^{-(c-c_0)/2}
\end{equation}

\noindent and so once $\langle N(c_0) \rangle$ is determined, the
p-value of a given observation and the corresponding significance
are readily obtained.

Naturally, one would like to choose the reference level $c_0$ in a
way that will minimize the resulting uncertainty on the bound. From
(\ref{eq3}), it can be seen that the statistical uncertainty is
proportional to $\sigma_N/\langle N \rangle$ where $\sigma_N$ is the
standard deviation of $N$. As long as the the distribution of $N$ is
`well behaved' in the sense that the relative uncertainty decreases
with increasing $\langle N \rangle$, then the optimal choice of
$c_0$ would be that which maximizes the expected number of
upcrossings (\ref{eq2}), that is $c_0=s-1$. Note that in the case
$s=1$, the maximal number of upcrossings occurs at $c_0 \to 0$,
however the ability to reliably estimate $\langle N(c_0) \rangle$ at
very low levels depends on the numerical resolution at which
$q(\theta)$ is calculated. In this case, one should therefore aim at
having $c_0$ as low as possible but still significantly larger than
the numerical resolution of $q(\theta)$, and the typical distance
between upcrossings should be kept significantly larger than the
$\theta$ resolution. In the example considered in the following
section $c_0=0.5$ satisfies those conditions and proves to be a good
choice.

It is further interesting to note that the dependence of $\langle
N(c) \rangle$ on $c$ in (\ref{eq2}), is the same as the asymptotic
form of the cumulative distribution of a $\chi^2$ variable with
$s+1$ degrees of freedom,

\begin{equation}\label{eq4}
P(\chi^2_{s+1} > c) \xrightarrow[c \to \infty]{}
\frac{c^{\frac{1}{2}(s+1)-1} e^{-c/2} }{
2^{\frac{1}{2}(s+1)-1}\mathrm\Gamma((s+1)/2) }
\end{equation}

allowing us to write, for large $c$ ($c \gg s$),
\begin{equation}\label{eq5}
P(q(\hat \theta) > c) \approx P(\chi^2_s > c) + \mathcal{N}
P(\chi^2_{s+1}
> c)
\end{equation}

where

\begin{equation}\label{eq6}
\mathcal{N} = \frac{1}{\sqrt{2\pi}}  \int_L^U C(\theta) d\theta
\end{equation}

And the bound has been replaced by a `$\approx$' sign since we are
dealing with the large $c$ limit. The probability described by eq.
(\ref{eq5}) has a natural interpretation. It is the same as one
would have for a random variable that is the maximal of $n$
independent $\chi^2_{s+1}$ variates and one $\chi^2_{s}$ variate,
with $E[n]=\mathcal{N}$. That is, if

\begin{equation}\label{eq7}
y = \max[x_0, x_1 ... x_n]
\end{equation}

and

\begin{center}
\begin{tabular}{ l c c r }
 $x_0 \sim \chi^2_{s}$, & $x_i \sim \chi^2_{s+1}$ &  $i=1...n$, &  $E[n] =
 \mathcal{N}$
\end{tabular}
\end{center}

then the tail probability $P(y>c)$ for large $c$ is given by the
right hand side of (\ref{eq5}).

Intuitively, this suggests that we can view the range of $\theta$ as
being composed of several (on average $\mathcal{N}$) independent
regions , where in each one the likelihood fit involves an extra
degree of freedom due to the variable mass, leading to a
$\chi^2_{s+1}$ distribution. The $\chi^2_s$ term accounts for the
possibility of having a local maximum on the edge of the allowed
region (assuming, from symmetry reasons, that each edge has a
probability $1/2$ of being a local maximum). The number
$\mathcal{N}$ can therefore be interpreted as an `effective number'
of independent search regions.

We remark that exactly such an intuitive reasoning was employed by
Quayle \cite{quayle} as a conjecture for the distribution of $q(\hat
\theta)$. It was found that the distribution of a random variable
defined according to (\ref{eq7}) reproduces, to a good
approximation, that of $q(\hat \theta)$, and that the agreement is
better at the tail. As shown above, this behavior is expected and
the conjecture is in fact a limiting case of Davies' formula
(\ref{eq1}).

\subsection{Trial factors}

It is sometimes useful to describe the look-elsewhere effect in
terms of a trial factor, which is the ratio between the probability
of observing the excess at some fixed mass point, to the probability
of observing it anywhere in the range. From (\ref{eq5}), we have

\begin{eqnarray}
\label{trial} trial\# & = & \frac{P(q(\hat \theta) > c)}{P(q(\theta)
> c)}
\\ & \approx & 1 + \mathcal{N} \frac{ P(\chi^2_{s+1} > c) }{P(\chi^2_{s}>
c) } \\ & \approx &
1+\mathcal{N}\sqrt{\frac{c}{2}}\frac{\mathrm\Gamma(s/2)}{\mathrm\Gamma((s+1)/2)}
\end{eqnarray}

For the case $s=1$, $\sqrt{c}$ is just the `fixed' significance,
that is the quantile of a standard gaussian corresponding to a
p-value of chi-square with $s$ degrees of freedom. This also holds
asymptotically for $s>1$, as $Z_{fix} = \sqrt{c} +
\mathcal{O}(s\frac{\log(c)}{\sqrt{c}})$. We therefore have, for $c
\gg s$,

\begin{equation}
\label{trial2} trial\#  \approx 1 + \frac{1}{\sqrt{2}}\mathcal{N}
Z_{fix} \frac{ \mathrm\Gamma(s/2)}{ \mathrm\Gamma((s+1)/2)}
\end{equation}

which, for the common case of $s=1$, is

\begin{equation}
\label{trial1} trial\#_{s=1}  \approx 1 + \sqrt{\frac{\pi}{2}}
\mathcal{N} Z_{fix}
\end{equation}

The trial factor is thus asymptotically linear with both the
effective number of independent regions, and with the fixed-mass
significance.

\section{Toy model simulations}\label{sec:toy}

\begin{figure*}[htbp] 
    \centering
    \includegraphics[width=4in]{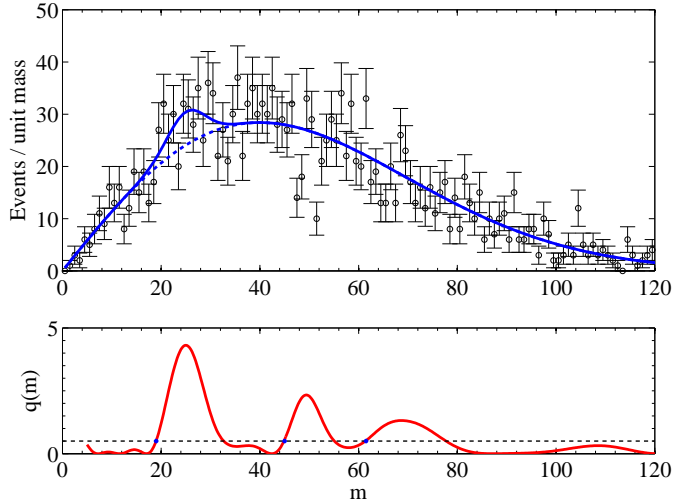}
    \caption{(top) An example pseudo-experiment with background only. The solid line shows the best signal fit, while the dotted line shows the background fit.
    (bottom) The likelihood ratio test statistic $q(m)$. The dotted line marks the reference level $c_0$ with the upcrossings marked by the dark dots.
    Note the broadening of the fluctuations as $m$ increases, reflecting the increase in the signal gaussian width. }
    \label{fig1}
 \end{figure*}

We shall now illustrate the procedure described above with a simple
example. Our toy model consists of a gaussian signal (`mass bump')
on top of a continuous background that follows a Rayleigh
distribution, in a mass range [0,120]. The width of the gaussian
increases linearly with the mass, representing a case where the mass
resolution changes with the mass.
%

We assume that the background shape is known but its normalization
is not, so that it is a free parameter in the fit (i.e. a nuisance
parameter), together with the signal location and normalization. We
use a binned profile likelihood ratio as our test statistic, where
the number of events in each bin is assumed to be Poisson
distributed with an expected value
\begin{equation}
E(n_i) = \mu s_i(m) + \beta b_i
\end{equation}
where $\mu$ is the signal strength parameter, $s_i(m)$ corresponds
to a gaussian located at a mass $m$, $\beta$ is the background
normalization and $b_i$ are fixed and given by the Rayleigh
distribution. For simplicity of notation we will use in the
following $\mathbf{s} = \{s_i\}$ and $\mathbf{b} = \{\beta b_i\}$.
The hypothesis that no signal exists, or equivalently that $\mu=0$,
will be referred to as the null hypothesis, $H_0$. $\hat \mu$ and
$\hat{\mathbf{b}}$ will denote maximum likelihood estimators while
$\hat{\hat{\mathbf{b}}}$ will denote the conditional maximum
likelihood estimator of the background normalization under the null
hypothesis.

The test statistic $q(m)$ is defined as:
\begin{equation} \label{logLfix}
   q(m) = -2ln\frac{\mathcal{L}(\mathbf{\hat{\hat{b}}})}{\mathcal{L}(\hat \mu \mathbf{s}(m)+\hat{\mathbf{b}})}.
 \end{equation}
where $\mathcal{L}$ is the likelihood function. An example
background-only pseudo-experiment is shown in Fig.\ref{fig1},
together with $q(m)$.

\begin{figure}[htbp] 
        \centering
         \includegraphics[width=2.5in]{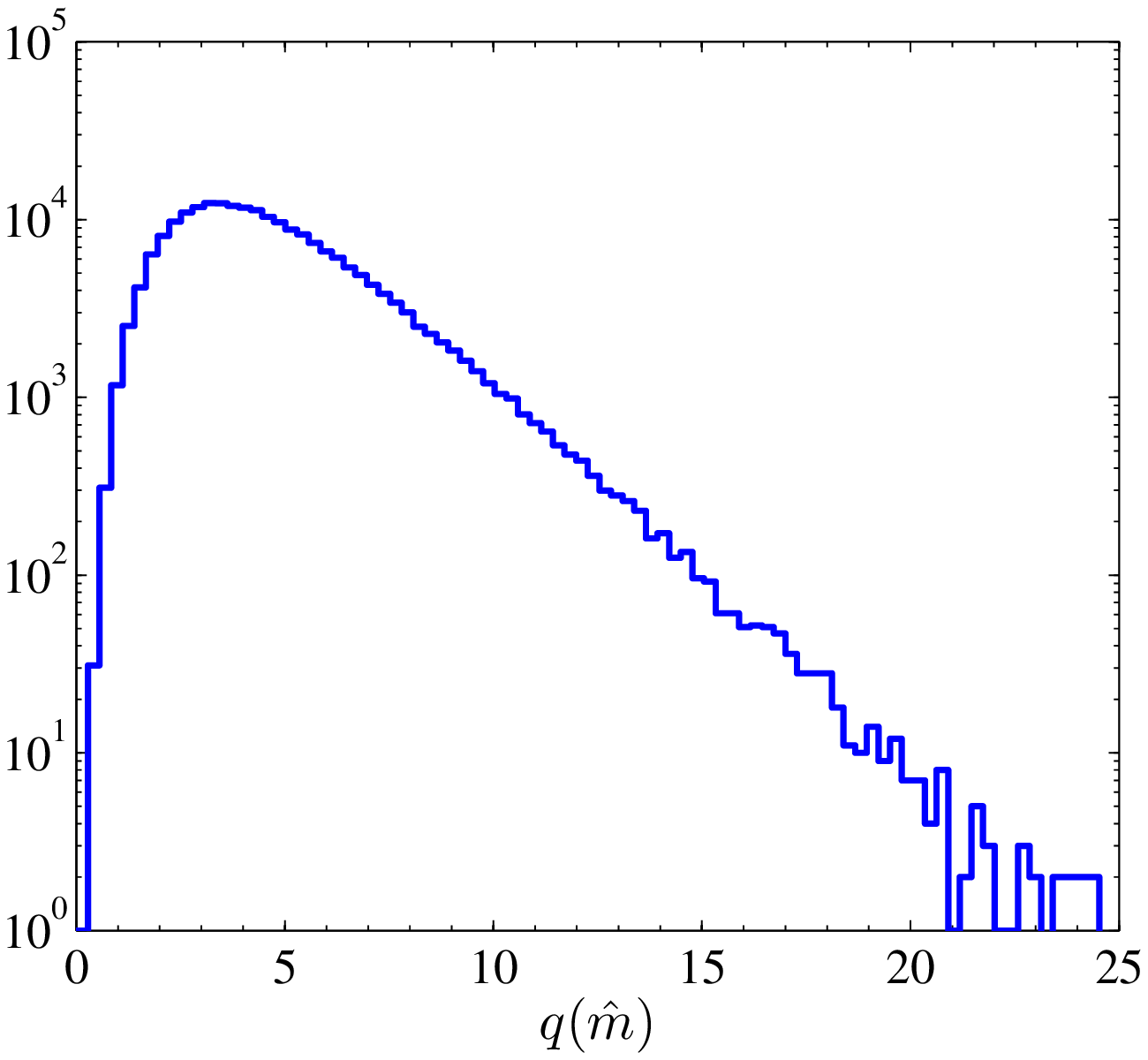}
        \includegraphics[width=2.5in]{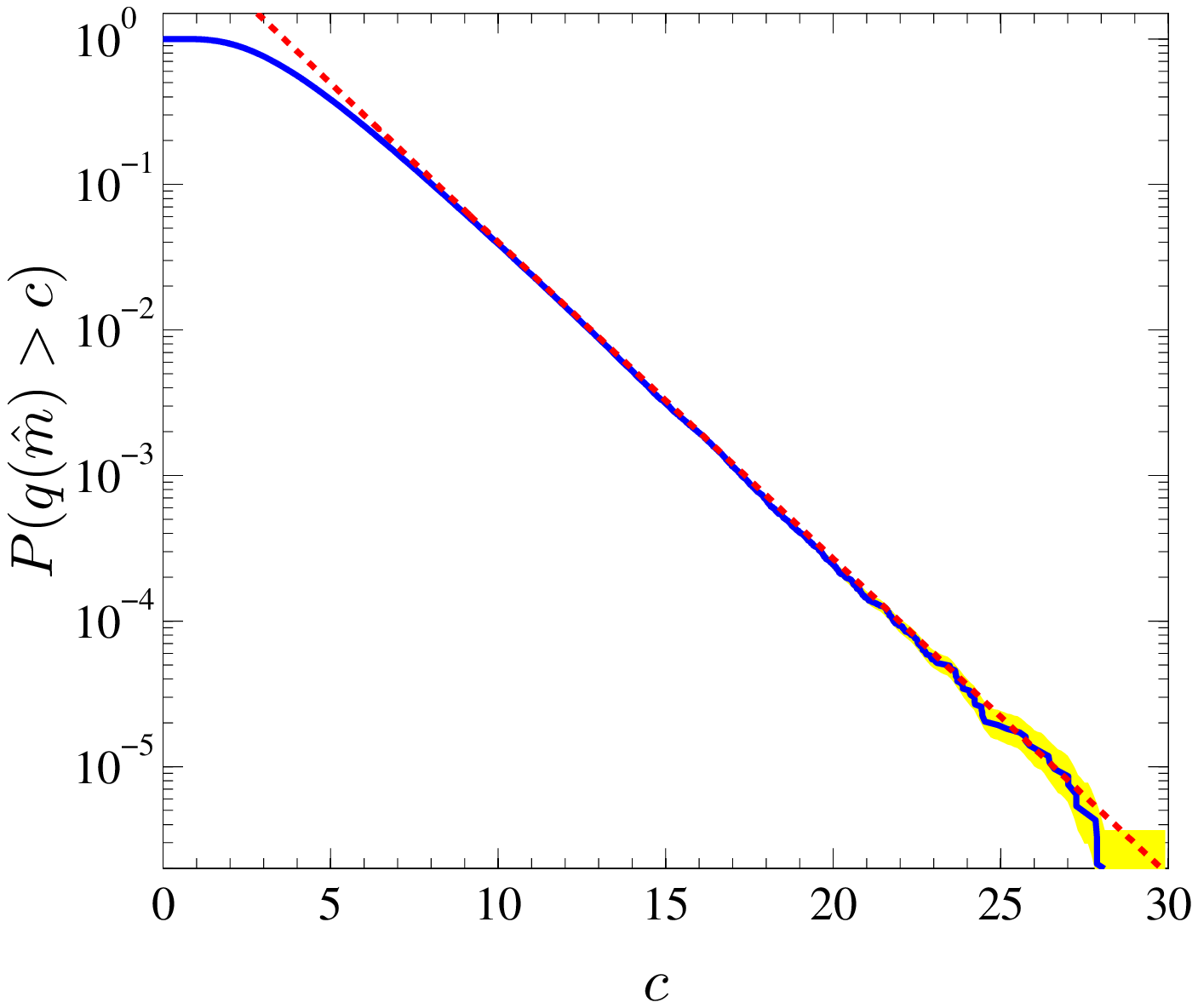}
        \caption{(top) Distribution of $q(\hat m)$. (bottom) Tail
        probability of $q(\hat m)$. The solid line shows the result
        of the Monte Carlo simulation, the dotted red line is the
        predicted bound (eq. \ref{eq3}) with the estimated $\langle N(c_0)
        \rangle$ (see text). The yellow band represents the statistical uncertainty due to the limited sample size.}
        \label{fig2}
\end{figure}

We choose a reference level of $c_0=0.5$ and estimate the expected
number of upcrossings, as demonstrated in Fig.\ref{fig1}, from a set
of 100 Monte Carlo simulations of background experiments. This gives
$\langle N(c_0) \rangle = 4.34 \pm 0.11$, which corresponds to
$\mathcal{N} = 5.58 \pm 0.14$. The distribution of $q(\hat m)$ is
then estimated from a sample of $\sim$1 million background
simulations and we compare the tail probabilities to the prediction
of (\ref{eq3}). The results are shown in Fig.\ref{fig2}. The bound
of (\ref{eq3}) gives an excellent approximation to the observed
p-values for large $c$.

\begin{figure}[hbp] 
   \centering
    \includegraphics[width=2.5in]{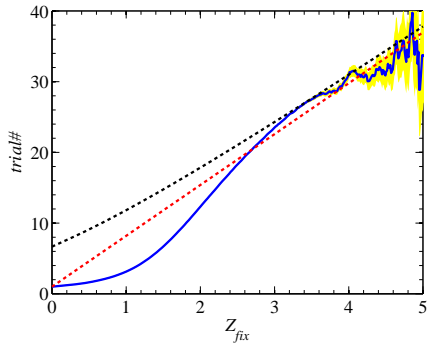}
    \caption{The trial factor estimated from toy Monte Carlo simulations (solid line), with the upper bound of
    eq.(\ref{eq3})
    (dotted black line) and the asymptotic approximation of eq.(\ref{trial1}) (dotted red line). The yellow band represents the statistical
uncertainty due to the limited sample size. }
    \label{fig3}
 \end{figure}

Figure \ref{fig3} shows the corresponding trial factor, compared to
the bound calculated from eq.(\ref{eq3})  and the asymptotic
approximation of eq.(\ref{trial1}).

\begin{figure}[htbp] 
        \centering
        \includegraphics[width=2.5in]{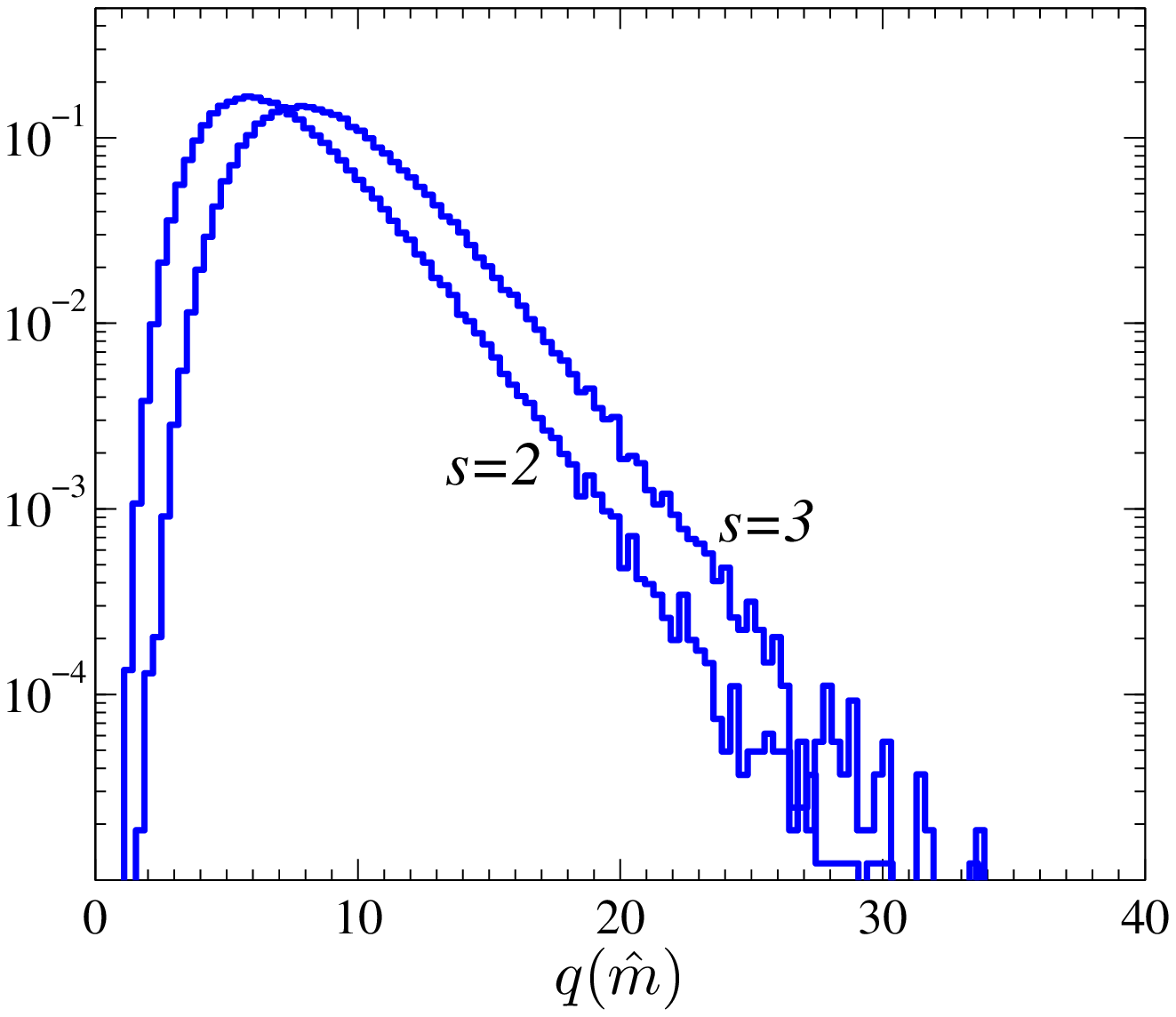}
        \includegraphics[width=2.5in]{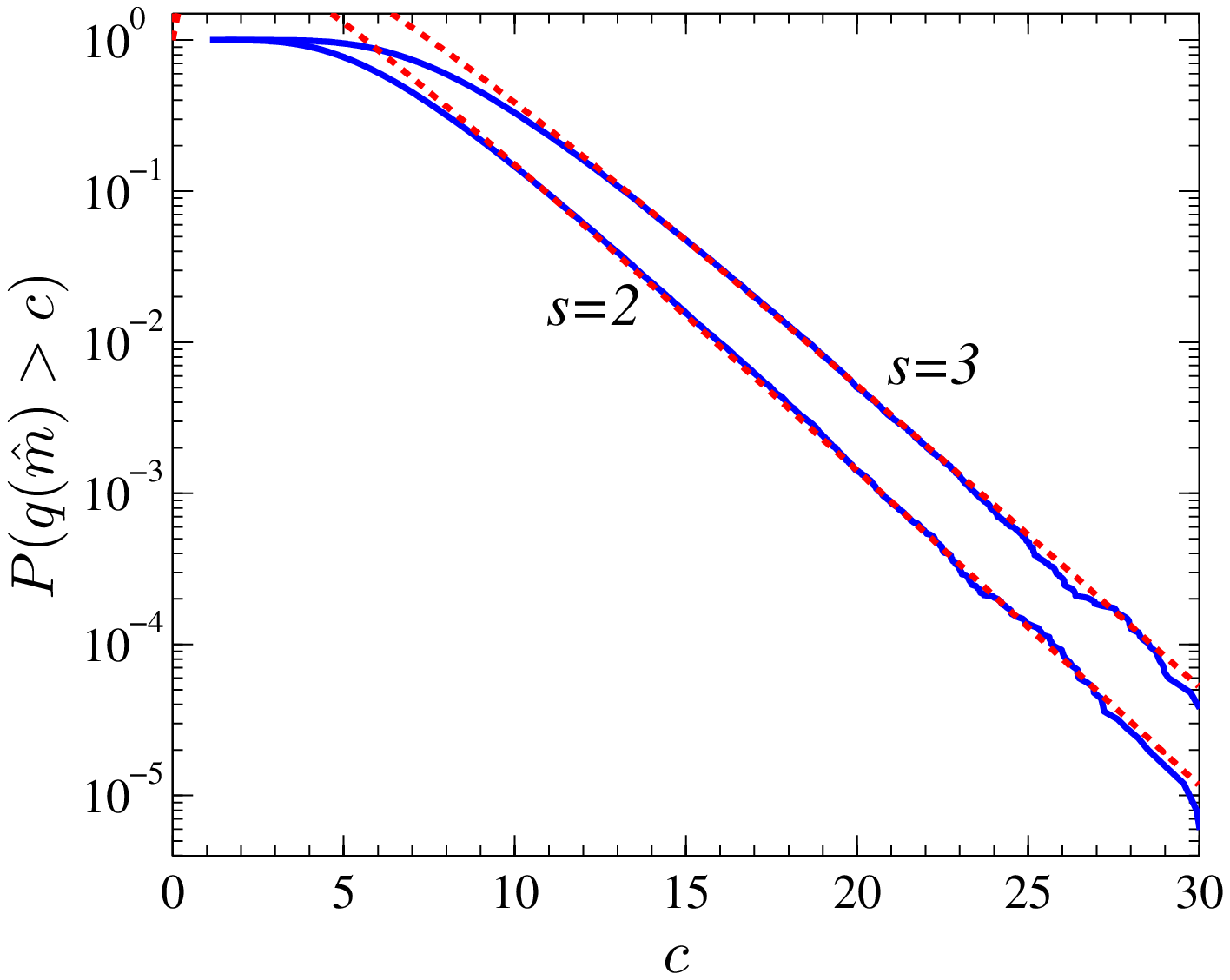}
        \caption{(top) Distribution of $q(\hat m)$ for $s=2,3$. (bottom)Tail
        probability of $q(\hat m)$. The solid lines shows the result
        of the Monte Carlo simulation, the dotted red lines are the
        predicted bound (eq. \ref{eq3}) with the estimated $\langle N(c_0)
        \rangle$ (see text).}
        \label{fig4}
\end{figure}

\begin{figure}[htbp] 
         \includegraphics[width=2.5in]{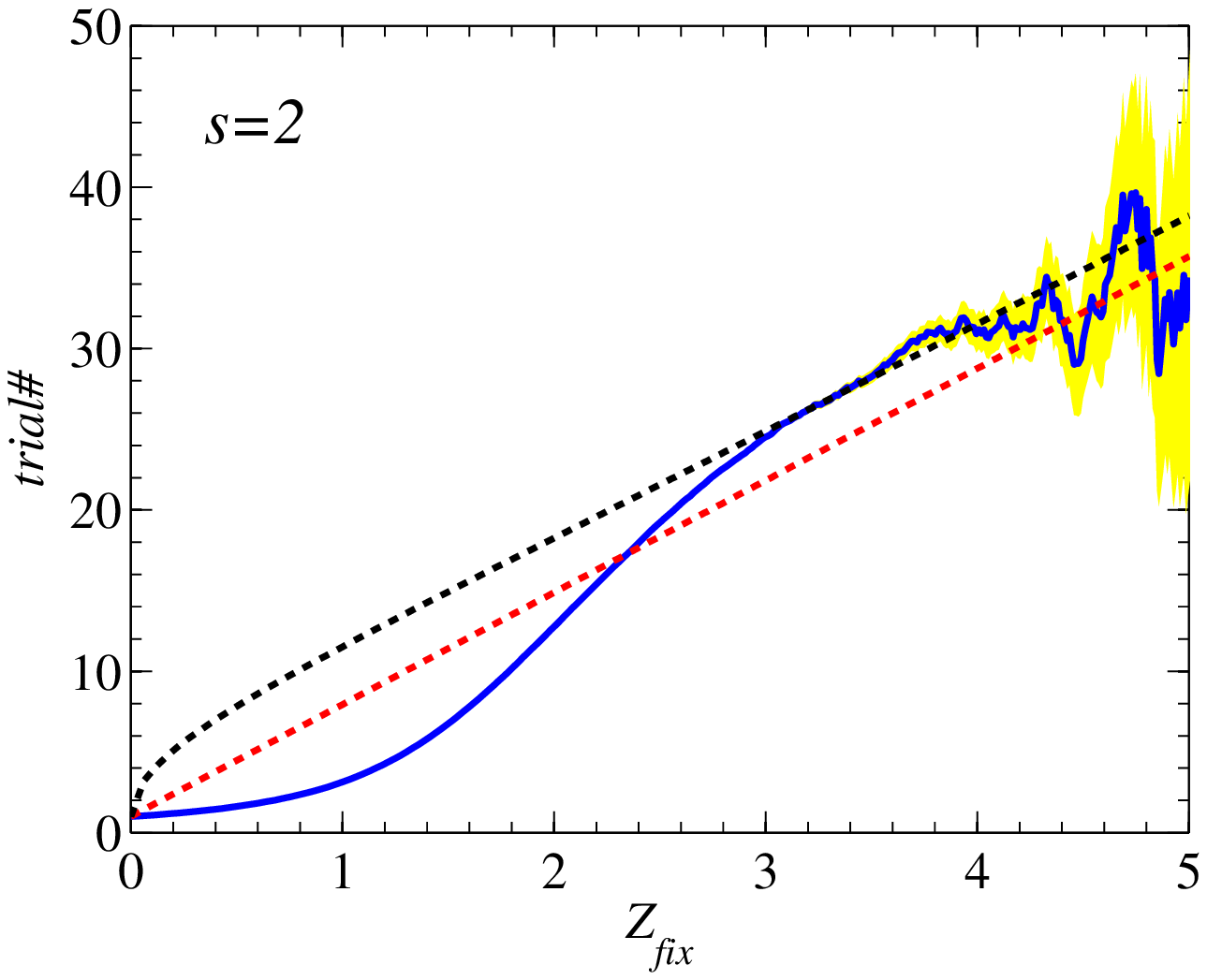}
        \includegraphics[width=2.5in]{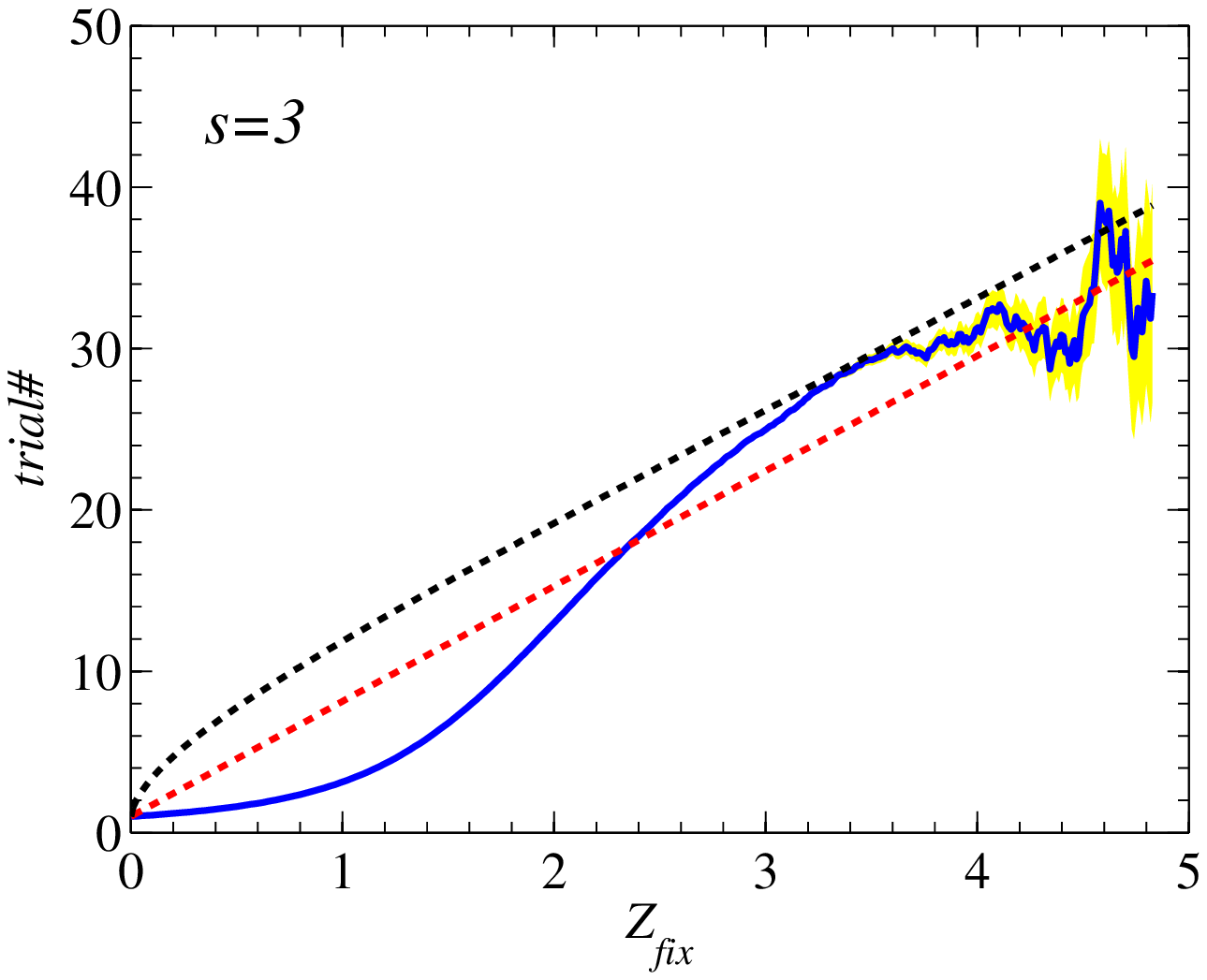}
        \caption{The trial factors estimated from toy Monte Carlo simulations (solid line), with the upper bound of
        eq.(\ref{eq3})
    (dotted black line) and the asymptotic approximation of eq.(\ref{trial2}) (dotted red line). The yellow band represents the statistical
uncertainty due to the limited sample size.}
        \label{fig5}
\end{figure}

We consider in addition a case where the number of degrees of
freedom is more than one. For this purpose, we assume several
independent channels, each identical to the one described above, and
where the signal normalizations $(\mu_1,...,\mu_s)$ are free
parameters. (This could represent, for example, a case where one is
searching for a resonance in several decay channels, with unknown
branching ratios). The reference level is chosen to be $c_0 = s-1$
as discussed in the previous section. The resulting distributions
and trial factors for $s=2,3$ are shown in figures \ref{fig4} and
\ref{fig5}. As before, the the bound (\ref{eq3}) agrees with the
observed p-value, within statistical variation. The rate at which
the asymptotic approximation (\ref{trial2}) converges to the bound
becomes slower when the number of degrees of freedom increases,
making it less accurate, however the trend of linear growth is
evident.

\section{Conclusions}

The look-elsewhere effect presents a case when the standard
regularity conditions of Wilks' theorem do not apply, and so
specialized methods are required for estimating tail probabilities
in the large sample limit. Such methods however exist, and as we
have demonstrated, can provide accurate results under fairly general
conditions. The procedure described in this paper consists of
estimating the expected number of upcrossings of the likelihood
ratio at a low reference level using a small set of Monte Carlo
simulations. This can then be related to the expected number of
upcrossings at a higher level using Davies' result (\ref{eq2}),
providing a bound on the probability of the likelihood ratio
exceeding this level, given by (\ref{eq3}). The method is easy to
implement in practical situations, and the bound converges to the
actual tail probability when this probability becomes small. It has
further been shown that the trial factor is asymptotically
proportional to the effective number of independent search regions
and to the fixed-mass significance, allowing for a simple
interpretation of the effect as being the result of two factors: the
first one is the mere fact that there are more available distinct
regions wherein fluctuations can occur, represented by the effective
number of independent regions; the second effect is that within each
region we further maximize the likelihood by fitting the mass in the
neighborhood of the fluctuation, which can be described by adding a
degree of freedom to the fit.

\section{Acknowledgments}
We are indebted to Michael Woodroofe, Luc Demortier and the referee
of this paper, for pointing us to the work of Davies \cite{davis87}
which became the leading thread of this work. We are grateful for
the valuable comments made by Luc Demortier in his summary talk in
the 2010 Banff workshop \cite{banff}, and to Michael Woodroofe who
spent valuable time in writing to us his impressions on the look
elsewhere effect. We also thank Bob Cousins for useful comments and
suggestions, BIRS and the organizers of the Banff workshop. One of
us (E.~G.) is obliged to the Benoziyo center for High Energy
Physics, to the Israeli Science Foundation (ISF), the Minerva
Gesellschaft, and the German Israeli Foundation (GIF) for supporting
this work.


\bibliographystyle{model1a-num-names}

\end{document}